\def\beq{\begin{equation}} 
\def\eeq{\end{equation}} 
\def\bed{\begin{displaymath}} 
\def\eed{\end{displaymath}} 
\def\beqq{\begin{eqnarray}} 
\def\eeqq{\end{eqnarray}} 
\def\bedd{\begin{eqnarray*}} 
\def\eedd{\end{eqnarray*}}

\def\bbb1{{\rm 1\!1}} 
 
\newcommand{\eqnl}[2]{\par\parbox{11cm} 
{\begin{eqnarray*}#1\end{eqnarray*}}\hfill 
\parbox{1cm}{\begin{eqnarray}\label{#2}\end{eqnarray}}\break}

\newcommand{\eqngrlb}[3]{\par\parbox{11cm} 
{\begin{eqnarray}\fbox{$\displaystyle#1\\#2$}\end{eqnarray}}\hfill 
\parbox{1cm}{\begin{eqnarray}\label{#3}\end{\eqnarray}}\break} 
 
\newcommand{\eqngrl}[3]{\par\parbox{11cm} 
{\begin{eqnarray*}#1\\#2\end{eqnarray*}}\hfill 
\parbox{1cm}{\begin{eqnarray}\label{#3}\end{eqnarray}}\break}

\newcommand{\eqngrrl}[4]{\par\parbox{11cm} 
{\begin{eqnarray*}#1\\#2\\#3\end{eqnarray*}}\hfill 
\parbox{1cm}{\begin{eqnarray}\label{#4}\end{eqnarray}}\break} 
 
\newcommand{\refs}[1]{(\ref{#1})}

\documentstyle[12pt]{article} 
 
\advance\hoffset by -.35in 
\begin{document} 
\bibliographystyle{unsrt} 
\global\parskip 6pt
 
\def\pr{\prime} 
\def\pa{\partial} 
\def\es{\!=\!} 
\def\ha{{1\over 2}} 
\def\>{\rangle} 
\def\<{\langle} 
\def\mtx#1{\quad\hbox{{#1}}\quad} 
\def\pan{\par\noindent} 
\def\lam{\lambda} 
\def\La{\Lambda} 
 
\def\A{{\cal A}} 
\def\G{\Gamma} 
\def\Ga{\Gamma} 
\def\F{{\cal F}} 
\def\J{{\cal J}} 
\def\M{{\cal M}} 
\def\R{{\cal R}} 
\def\W{{\cal W}} 
\def\tr{\hbox{tr}} 
\def\al{\alpha} 
\def\d{\hbox{d}} 
\def\De{\Delta} 
\def\L{{\cal L}} 
\def\H{{\cal H}} 
\def\Tr{\hbox{Tr}} 
\def\I{\hbox{Im}} 
\def\R{\hbox{Re}} 
\def\ti{\int\d^2\theta} 
\def\bti{\int\d^2\bar\theta} 
\def\ttbi{\int\d^2\theta\d^2\bar\theta} 
 
 
\begin{titlepage}
\hfill{DIAS-STP-97-10}
\vspace*{2cm}
\begin{center}
{\Large\bf Three-Dimensional Black Holes and String Theory}\\
\vspace*{2cm}
Danny Birmingham\footnote{Email: dannyb@ollamh.ucd.ie}\\
{\em University College Dublin, Department of Mathematical Physics\\
Belfield, Dublin 4, Ireland}\\
\vspace{1cm}
Ivo Sachs\footnote{Email: ivo@stp.dias.ie}\\
{\em Dublin Institute for Advanced Studies, School of Theoretical Physics,\\
10 Burlington Road, Dublin 4, Ireland}\\
\vspace*{1cm}
Siddhartha Sen\footnote{Email: sen@maths.tcd.ie}\\
{\em Trinity College Dublin, Department of Mathematics,\\
Dublin 2, Ireland}
\vspace{2cm}
\begin{abstract}
The exact decay rate for emission of massless minimally coupled scalar 
fields by a non-extremal black hole in $2+1$ dimensions is obtained. 
At low energy, the decay rate into 
scalars with zero angular momentum is correctly reproduced within 
conformal field theory. The conformal field theory 
has both left- 
and right-moving sectors and their contribution to the decay rate
is associated naturally with left and right temperatures of the 
black hole.

\end{abstract}
\vspace{1cm}
July 1997
\end{center}
\end{titlepage}

\section{Introduction}
Recently, it has become clear that a large class of 
extremal and near-extremal black holes allow for a conformal field theory 
or effective string theory description. Extremal black holes often correspond 
to $B\!P\!S$-states of an underlying fundamental string theory. 
Agreement  between the Bekenstein-Hawking entropy 
and the counting of string states for extremal black 
holes in five dimensions was first obtained in \cite{SV}. However,  
the correspondence does not seem 
to be restricted to extremal black holes. Indeed, the entropy of near-extremal 
black holes is often completely described by an effective string theory 
\cite{Horowitz}-\cite{Callan96a}.  

On another front, the decay rates of non-extremal black holes have also been 
examined. This involves studying the absorption 
of quanta by the black hole,  and then 
allowing it to evaporate, via Hawking radiation, back to extremality. In 
\cite{Callan96a}-\cite{GK}, the low energy 
scattering cross sections and decay rates for a massless 
minimally coupled scalar field were computed for a large class of 
four- and five-dimensional black holes, 
and agreement was found with conformal field theory or effective string 
theory predictions. In each of these cases, the result 
relied on a particular matching
of solutions,  in a region near the black hole
horizon and an asymptotic region far from the black hole.
For certain ranges of parameters inherent to the problem, this matching 
agrees with a conformal field theory description. 

In this paper, we study the propagation a massless minimally coupled scalar 
field in the background geometry
of the $(2+1)$-dimensional  Ba\~{n}ados-Teitelboim-Zanelli black hole 
\cite{BTZ}.
This black hole is described by two parameters, its mass $M$  and angular
momentum $J$. In addition, the metric has constant negative curvature,
and is thus locally isometric to anti-de Sitter space. 
The special feature here is that the wave equation can be solved exactly, 
without any approximations \cite{IS}.
This allows us to determine exactly the range of energy and 
angular momentum of the scattered field, for which the the 
decay rate agrees with the conformal field theory description. We find 
agreement for energies small in comparison to the size of the black
hole, and to the curvature scale of the spacetime; in addition, one
is restricted to  the zero angular momentum wave. In this region, 
however, agreement is found for all values of $M$ and $J$, and thus the 
conformal field theory description is not restricted to a near-extremal limit.
Apart from that, we find behaviour similar to that observed in
five dimensions, namely that the conformal
field theory has both left- and right-moving sectors. The corresponding
decay rate is then written naturally in terms of left and right 
temperatures of the black hole. 

Similarly to the $5$-D black holes, the $BTZ$ black hole is a solution 
of string theory \cite{Kaloper,HorRevLett}.
The string scattering off $BTZ$ black holes has been
considered in \cite{Larsen,Sato} (see also \cite{Hyun}). 

\section{The BTZ Black Hole}  
Geometrically, three-dimensional anti-de Sitter space can be
represented as the ${S\!L(2,\mbox{\bf R})}$
group space. Isometries are then represented by elements
of the group  
$S\!L(2,\mbox{\bf R})\times S\!L(2,\mbox{\bf R})/\mbox{\bf Z}_{2}$,
where the two copies of ${S\!L(2,\mbox{\bf R})}$
act by left and right multiplication.
The $BTZ$ black hole is obtained as the quotient space
${S\!L(2,\mbox{\bf R})}/
\<(\rho_L,\rho_R)\>$, where  
$\<(\rho_L,\rho_R)\>$ denotes a certain finite subgroup of  
$S\!L(2,\mbox{\bf R})\times S\!L(2,\mbox{\bf R})/\mbox{\bf Z}_{2}$
generated by $(\rho_L,\rho_R)$ \cite{BHTZ}.  
We choose Schwarzschild-like coordinates in which the $BTZ$ metric reads 
\cite{BTZ,Carlip}  
\eqnl{ds^2 = -( N^\perp)^2dt^2 + f^{-2}dr^2 
  + r^2\left( d\phi + N^\phi dt\right)^2,}{m1} 
with lapse and shift functions and radial metric 
\eqnl{ 
N^\perp = f 
  = \left( -M + {r^2\over \ell^2} + {J^2\over 4r^2} \right)^{1/2}, 
  \;\; N^\phi = - {J\over 2r^2}, \;\;  (|J|\le M\ell).}{a2} 
The metric \refs{m1} is singular when $r\!=\!r_\pm$, where 
\eqnl{ 
r_\pm^2={M\ell^2\over 2}\left \{ 1 \pm 
\left [ 1 - \left({J\over M\ell}\right )^2\right ]^{1/2}\right \},}{a6} 
i.e., 
\eqnl{ 
M={r_+^2+r_-^2\over \ell^2},\;\; J={2r_+ r_-\over \ell}.}{a6a} 

The $M\!=\!-1$, $J\!=\!0$ metric may be recognized as that of ordinary 
anti-de Sitter space; it is separated by a mass gap from the $M\!=\!0$, 
$J\!=\!0$ ``massless black hole", whose geometry is discussed in 
Refs.~\cite{BHTZ} and \cite{Steif2}. 
For convenience, we recall that the Hawking temperature $T_{H}$,
the area of the event horizon $\A_{H}$, and the angular 
velocity at the event horizon $\Omega_{H}$,
are given by
\eqnl{
T_{H} = \frac{r_{+}^{2} - r_{-}^{2}}{2 \pi \ell^{2}r_{+}},\;\;
\A_{H} = 2 \pi r_{+},\;\; \Omega_{H} = \frac{J}{2 r_{+}^{2}}.}{temp}

The $BTZ$ black hole is part of a solution of low energy string theory 
\cite{Kaloper,HorRevLett}. The low energy string effective action is
\beq
I = \int\!d^3x\,\sqrt{-g}e^{-2\phi}\left( {4\over k} + R
  + 4\nabla_\mu\phi\nabla^\mu\phi
  - {1\over12}H_{\mu\nu\rho}H^{\mu\nu\rho} \right) ,
\label{h4}
\eeq
where $\phi$ is the dilaton, $H_{\mu\nu\rho}$ is an antisymmetric
Kalb-Ramond field, which in three dimensions must be proportional to the
volume form $\epsilon_{\mu\nu\rho}$, and $k$ is the cosmological constant.
It was observed in \cite{Kaloper,HorRevLett}
that, in three dimensions, the ansatz
\beq
H_{\mu\nu\rho} = {2\over \ell}\epsilon_{\mu\nu\rho}, \quad \phi=0 ,
\quad k=\ell^2,
\label{h5}
\eeq
reduces the equations of motion of  \refs{h4} to the Einstein
field equations satisfied by \refs{m1}. 
In fact, there is a corresponding exact solution of string
theory, the $\hbox{SL}
(2,\mbox{\bf R})$
WZW model with an appropriately chosen central charge describes
the propagation of strings. By quotienting out the discrete group
$\langle(\rho_L,\rho_R)\rangle$ by means of an
orbifold construction, one obtains a theory that may be shown to be
an exact string theoretical representation of the $BTZ$ black hole
\cite{Kaloper,HorRevLett}.

\section{The Wave Equation} 
It has been known for some time that the minimally coupled scalar field 
equation can be solved exactly in the background
geometry of the $BTZ$ black hole \cite{IS}. This will allow us to determine 
the scattering cross section and decay rate of the scalar field exactly.  
Substitution of the metric \refs{m1} into the covariant Laplacian 
\eqnl{\Box=\frac{1}{\sqrt{|g|}}\pa_\mu\sqrt{|g|}\;g^{\mu\nu}\pa_\nu}{box1} 
leads to the scalar wave equation 
\eqnl{\left(-f^{-2}\pa_t^2+f^2\pa_r^2+\frac{1}{r}
\left(\pa_r rf^2\right)\pa_r- 
\frac{J}{r^2}f^{-2}\pa_t\pa_\phi-\frac{A}{r^2}f^{-2}\pa_\phi^2\right)
\Psi=0,}{box2} 
where
\eqnl{f^2=\frac{1}{\ell^2 r^2}(r^2-r_-^2)(r^2-r_+^2),\;\; 
A=M-\frac{r^2}{\ell^2}.}{c1}
This suggests the ansatz 
\eqnl{\Psi(r,t,\phi)=R(r,\omega,m)e^{-i\omega t+im\phi},}{A1} 
leading to the radial equation for $R(r)$ 
\begin{eqnarray}
\pa_r^2R(r)&+&\left(-\frac{1}{r}+\frac{2r} 
{r^2-r_-^2}+\frac{2r}{r^2-r_+^2}\right)\partial_{r}R(r)
+f^{-4}\left(\omega^2-\frac{J\omega m} 
{r^2}+\frac{Am^2}{r^2}\right)R(r).\nonumber\\
\label{box5}
\end{eqnarray} 
Changing variables to $x=r^2$, the radial equation becomes 
\eqnl{(x-x_-)(x-x_+)\pa_x^2\;R(x)+[2x-x_+ - x_-]\pa_x\;R(x)+K(x)\;R(x)=0,} 
{box6} 
where 
\eqnl{K(x)=\frac{\ell^2}{4f^2}\left(\omega^2-\frac{J\omega m}{x} 
+\frac{Am^2}{x}\right).} 
{box7} 

We introduce a further change of variables by defining 
\eqnl{z=\frac{x-x_+}{x-x_-}.}{v1} 
The radial equation then becomes
\eqnl{z(1-z)\pa_z^2\;R(z)+(1-z)\pa_z\;R(z)  
+\left( \frac{A_{1}}{z} + B_{1}\right) R(z)=0,}{box8} 
where  
\eqnl{A_{1}=\left(\frac{\omega-m\Omega_H}{4\pi T_H}\right)^2,\;\; 
B_{1}=-\frac{x_-}{x_+}\left(\frac{\omega-m\Omega_H\frac{x_+}{x_-}}{4\pi T_H} 
\right)^2.}{AB} 
The hypergeometric form of \refs{box8} becomes explicit upon removing the pole in the last term through the ansatz 
\eqnl{R(z)=z^\al g(z),\;\;\alpha^{2} =-A_{1}.}{A2} 
We then have
\eqnl{z(1-z)\pa_z^2g(z)+(2\alpha+1)(1-z)\pa_zg(z)+(A_{1} +B_{1}) 
g(z)=0.}{aldet} 
In the neighbourhood of the horizon, $z\es0$,
two linearly independent solutions are then given by 
$F(a,b,c,z)$ and $z^{1-c}F(a-c+1,b-c+1,2-c,z)$, where
\eqngrrl{a+b&=&2\al,} 
        {ab&=&\al^2-B_{1},} 
        {c&=&1+2\al.}{abc} 
Note that $c = a+b+1$. 

\section{The Decay Rate}
We choose the solution which has ingoing flux at the horizon, namely,
\eqnl{R(z)=z^\al F(a,b,c,z).}{sol1} 
To see this, we note that the conserved flux for \refs{box5} is given, 
up to an irrelevant  
normalisation, by  
\eqnl{\F=\frac{2\pi}{i}\left(R^*\Delta\pa_r R-R\Delta\pa_r R^*\right),}{flux1} 
where $\Delta= r f^2$. The flux can be evaluated by noting that
\eqnl{\Delta\pa_r=\frac{2\Delta_-}{\ell^2}z\pa_z,}{dz} 
where $\Delta_{-} = x_{+} - x_{-}$. 
Then, using  the fact that $ab$ is real, 
we find the total flux (which is independent of $z$) to be given by 
\eqnl{\F(0)=\frac{8\pi\Delta_-}{\ell^2}\I[\al]\left|F(a,b,c,0)\right|^2=2{\cal{A}}_H(\omega-m\Omega_H).}{f2} 

In order to compute the absorption cross section, we need to divide \refs{f2}  
by the ingoing flux at infinity. The singularity of \refs{box5} at infinity  
is such that it admits one solution of the form \cite{Rab}
\eqnl{u_1(y)=y^2\sum\limits_{n= 0}^\infty c_n y^n,}{se1} 
where $y\es \ell/r$.  
A second linearly independent solution of \refs{box5} at 
infinity is then given by  
\eqnl{ u_{2}(y) = \sum\limits_{n= 0}^\infty d_n y^n +A u_1(y)\log(y),}{se2} 
where $A$ is a constant.
Up to second order in $y$, we have 
\eqnl{u_1(y)=y^2,\;\;u_2(y)=1
-\left(\frac{\omega^2\ell^2-m^2}{2}\right)y^2\log(y).}{se12} 
The distinction between ingoing and outgoing waves is complicated by the fact
that the $BTZ$-spacetime is not asymptotically flat.
For a tachyon field, ingoing
and outgoing waves have been defined in e.g. \cite{Sato}.
A naive extrapolation to massless fields is not sensible as the resulting
ingoing and outgoing waves are given by $u_1$ and $u_2$ in \refs{se12},
which both
have vanishing flux. However, we can define ingoing 
and outgoing waves
to be complex linear combinations
of $u_{1}$ and $u_{2}$ which have positive
and negative flux, respectively. This leads to
\eqnl{R^{in}=A_i\left(1-i\frac{c\ell^2}{r^2}\right),\;\; R^{out}= 
A_o\left(1+i\frac{c\ell^2}{r^2}\right),}{inout} 
where $c$ is some positive dimensionless constant, which we take to be 
independent of  the frequency $\omega$.  
We note the comparison here with the near region behaviour of
the ingoing and outgoing solutions in the four-dimensional case 
(eqn. (2.22) in 
\cite{Strominger1}).
The ingoing flux is correspondingly 
\eqnl{\F_{in}=8 \pi c|A_i|^2.}{fi} 

The asymptotic behaviour of \refs{sol1} for large $r$ is readily available
\cite{Abramowitz}, and we can then match this to \refs{inout} to
determine the coefficients $A_{i}$ and $A_{o}$.  We find
\begin{eqnarray}
A_i+A_o&=& \frac{\Gamma(a+b+1)}{\Gamma(a+1)\Gamma(b+1)},\nonumber \\ 
A_i-A_o&=&-\frac{\Delta_-\Gamma(a+b+1)}{ic\ell^2}
\left\{\frac{\log(\Delta_-/\ell^2)+\psi(a+1)+\psi(b+1) 
-\psi(1)-\psi(2)}{\Gamma(a)\Gamma(b)}\right.\nonumber \\
&+&\left.\frac{\al}{\Gamma(a+1)\Gamma(b+1)}\right\},
\label{match} 
\end{eqnarray}
where $\psi$ is the digamma function.
We can estimate the relative importance of the two terms in \refs{match}
as follows. Firstly, we note that   
\eqnl{ab= -\frac{\ell^2}{4 \Delta_-}(\omega^2 \ell^2 - m^2),\;\; \alpha=i 
\frac{(\omega-m\Omega_H) \ell^2r_+}{2\Delta_-}.}{us1} 
Using $\Gamma(z +1) = z \Gamma(z)$, we find
\begin{eqnarray} 
A_i-A_o&=&-\frac{\Gamma(a+b+1)}{c\Gamma(a+1)\Gamma(b+1)} 
\Big\{\frac{(\omega -m\Omega_H)r_+}{2} +\frac{i(\omega^2 \ell^2-m^2)}{4}
\Big[\log(\Delta_-/\ell^2)\nonumber\\ 
&+&\psi(a+1)+ 
\psi(b+1)-\psi(1)-\psi(2)\Big]\Big\}.
\label{match2} 
\end{eqnarray} 
Furthermore, 
\eqnl{a=\frac{i\ell}{2(r_+-r_-)}\left(\ell\omega-m\right),\;\; 
b=\frac{i\ell}{2(r_++r_-)}\left(\ell\omega+m\right).}{match3} 
If $m\es0$ and $\omega\!<\!<\!\hbox{min}(\frac{1}{r_+},\frac{1}{\ell})$, the  
difference $A_{i} - A_{o}$  in \refs{match2} is small compared to the 
sum $A_{i} + A_{o}$, so that
\eqnl{A_i\simeq \frac{1}{2}\frac{\Gamma(a+b+1)}{\Gamma(a+1) 
\Gamma(b+1)}.}{match4}
This approximation means that the Compton wavelength of the scattered 
particle is much
bigger that the size of the black hole and the scale set by the curvature
of the anti-de Sitter space.
Note that the logarithmic term in \refs{match2} ensures that the term
in braces is finite for all values of $\ell/r_+$, i.e., all values of $M$ 
and $J$. In particular, the extreme limit $J = \pm M\ell$, or
equivalently $\Delta_{-} = 0$,
is well defined. 
Hence, for $m\es0$ and $\omega$ small in the sense defined above, the 
approximation \refs{match4} should be valid. 
For $m \neq 0$ on the other hand, 
there is no 
obvious choice of black hole parameters for 
which the term in braces in \refs{match2} is small, and  hence  
\refs{match4} is modified for $m\neq 0$.  
 
Let us  consider the  
$m\es 0$ wave and assume 
$\omega\!<\!<\!\hbox{min}(\frac{1}{r_+},\frac{1}{\ell})$  
so that \refs{match4} is valid. Then the partial wave absorption cross 
section  
is given by  
\eqnl{\sigma^{m=0}=\frac{\F(0)}{\F_{in}}=\frac{1}{\pi c}{\cal{A}}_H
\omega\frac{| 
\Gamma(a+1)\Gamma(b+1)|^2}{|\Gamma(a+b+1)|^2}.}{sigma1} 
In order to relate the partial wave cross section to the plane wave cross  
section $\sigma_{abs}$, we need to divide $\sigma^{m=0}$ by $\omega$  
\cite{Gibbons}. We find  
\eqnl{\sigma_{abs}= {\cal{A}}_H\frac{| 
\Gamma(a+1)\Gamma(b+1)|^2}{|\Gamma(a+b+1)|^2},}{sigma2} 
where we have chosen $c$ so that $\sigma_{abs}(\omega)
\rightarrow{\cal{A}}_H $  
for $\omega\rightarrow 0$ \cite{Gibbons}. The decay rate $\Gamma$ of a 
non-extremal black hole is then given by 
\eqngrl{\Gamma&=&\frac{\sigma_{abs}}{e^{\frac{\omega}{T_H}}-1}=
T_H{\cal{A}}_H\omega^{-1}e^{-\frac{\omega}{2T_H}}| 
\Gamma(a+1)\Gamma(b+1)|^2}
{&=&4\pi^2\ell^2\omega^{-1}T_LT_Re^{-\frac{\omega}{2T_H}}\left| 
\Gamma\left(1+i\frac{\omega}{4\pi T_L}\right)\Gamma\left(1+i
\frac{\omega}{4\pi T_R}\right)\right|^2,}{decay1}
where the left and right temperatures are defined by  
\eqnl{T_{L/R}^{-1}=T_H^{-1}\left(1\pm\frac{r_-}{r_+}\right),}{TLR}
and we have used ${\cal{A}}_H=2\pi r_+$. 

We now compare this decay rate with the conformal field theory prediction. 
As explained in \cite{mathur,Strominger1,gubser}, in the effective string 
theory picture such decays are described by a coupling of the spacetime 
scalar field to an operator with dimension $1$ in the conformal field 
theory, both in the left- and right-moving sector. This might have been 
expected since the WZW-model of which the 
$BTZ$-black hole is a classical solution contains both chiralities 
\cite{HorRevLett}. A calculation closely analogous to that 
presented in \cite{Strominger1} shows the emission rate is given by 
\eqnl{\int d\sigma^+\;e^{-i\omega(\sigma^+-i\epsilon)}\left[
\frac{2 T_R}{\sinh(2 \pi T_R\sigma^+)}\right]^{2}
\int  d\sigma^-e^{-i\omega(\sigma^--i\epsilon)}\left[
\frac{2 T_R}{\sinh(2 \pi T_R\sigma^-)}\right]^{2}.}{CFT1}
Performing the $\sigma^\pm$ integrals 
we then obtain, up to an undetermined numerical constant,
\eqnl{\Gamma=4\ell^2\omega^{-1}T_LT_Re^{-\frac{\omega}{2T_H}}\left| 
\Gamma\left(1+i\frac{\omega}{4\pi T_L}\right)\Gamma\left(1+
i\frac{\omega}{4\pi T_R}\right)\right|^2,}{decay2}
where $\ell^2$ has been included for dimensional reasons, and a 
factor $1/\omega$ 
accounts for the normalisation of the outgoing 
scalar \cite{Strominger1}. Hence, the $CFT$-prediction \refs{decay2} and the 
semiclassical decay rate \refs{decay1} agree. 
Note that the above analysis is valid for all values of $M$ and $J$, 
subject to the low energy restriction.  
Thus, for the three-dimensional black hole the $CFT$ description is 
not restricted to the near-extreme,
or near-$B\!P\!S$ limit; we recall that an extreme $BTZ$-black hole is also
a $B\!P\!S$ configuration, as shown in \cite{CH}. 

Note also that supersymmetry has not played a role in the 
calculations presented here. 
In the four-dimensional case 
\cite{Strominger1}, supersymmetry appeared indirectly in determining the 
relevant conformal field theory. However, such an analysis 
fails in the $3$-dimensional case. Similarly, in five dimensions, bosons and 
fermions must propagate on the effective string in order to recover the decay 
rates for even and odd angular momenta alike. Again, this fails in 
three dimensions because of the restriction $m\es 0$. 
On the other hand, it has been shown \cite{CH} that the $BTZ$-black hole 
admits one supersymmetry in the extremal case with $M\neq0$, and two 
supersymmetries if furthermore $M\es0$. In the latter  case,
it may be viewed as the ground state of $(1,1)$-adS supergravity. 

A final remark concerns the duality between the $BTZ$-black hole and the 
black string \cite{HorRevLett}. This suggests that an
analysis similar to that 
presented here might go through for the black string. A question that arises 
naturally here is how the two low energy decay rates compare. 
For the tachyon, the reflection coefficients have been obtained in
\cite{Sfetsos} for the black string, and in
\cite{Larsen,Sato} for the case
of the black hole.

\vspace{1cm}
\noindent{\large \bf Acknowledgements}\\
D.B. would like to thank the Theory Division at CERN for hospitality
during the completion of this work.

\end{document}